\documentclass[reprint,superscriptaddress,twocolumn,amsmath,amssymb,]{revtex4-2}
\pdfoutput=1
\usepackage{footmisc}
\usepackage{xcolor}
\usepackage{graphicx}% Include figure files
\usepackage{dcolumn}% Align table columns on decimal point
\usepackage{bm}% bold math
\usepackage{romannum}
\usepackage{siunitx}
\usepackage{epstopdf}
\usepackage{titlesec}
\titleformat{\section}{\raggedright\fontsize{12.5}{25}\bfseries}{\arabic{section}.}{1em}{}

\DeclareUnicodeCharacter{0301}{\'a}
\DeclareUnicodeCharacter{00D6}{\"o}
\DeclareUnicodeCharacter{00C4}{\"a}
\DeclareUnicodeCharacter{00FC}{\"u}
\DeclareUnicodeCharacter{00DF}{\ss}

\usepackage{hyperref}
\usepackage[normalem]{ulem}

\begin{document}

\pagenumbering{arabic}

\title{\fontsize{15}{19}\selectfont Magneto-optical study of metamagnetic transitions in the antiferromagnetic phase of $\alpha$-RuCl$_3$}

\author{Julian Wagner}

\author{Anuja Sahasrabudhe}
\affiliation{Universit\"at zu K\"oln, II. Physikalisches Institut, Z\"ulpicher Stra\ss e 77, K\"oln D-50937, Germany}

\author{Rolf Versteeg}

\affiliation{Laboratoire de Spectroscopie Ultrarapide and Lausanne Centre for Ultrafast Science (LACUS),\\ ISIC-FSB, \'Ecole Polytechnique F\'ed\'erale de Lausanne, CH-1015 Lausanne, Switzerland}

\author{Lena Wysocki}
\affiliation{Universit\"at zu K\"oln, II. Physikalisches Institut, Z\"ulpicher Stra\ss e 77, K\"oln D-50937, Germany}

\author{Zhe Wang}
\affiliation{Department of Physics, TU Dortmund University, Otto-Hahn-Str. 4, 44227 Dortmund, Germany}

\author{V.~Tsurkan}
\affiliation{Experimental Physics V, Center for Electronic Correlations and Magnetism, University of Augsburg, 86135 Augsburg, Germany}
\affiliation{ Institute of Applied Physics, MD 2028, Chisinau, Republic of Moldova}

\author{A.~Loidl}
\affiliation{Experimental Physics V, Center for Electronic Correlations and Magnetism, University of Augsburg, 86135 Augsburg, Germany}

\author{D.~I.~Khomskii}

\author{Hamoon Hedayat}
\email{hedayat@ph2.uni-koeln.de}
\author{Paul H. M. van Loosdrecht}
\email{pvl@ph2.uni-koeln.de}
\affiliation{Universit\"at zu K\"oln, II. Physikalisches Institut, Z\"ulpicher Stra\ss e 77, K\"oln D-50937, Germany}

\begin{abstract}

$\alpha$-RuCl$_3$ is a promising candidate material to realize the so far elusive quantum spin liquid ground state. However, at low temperatures, the coexistence of different exchange interactions couple the effective pseudospins into an antiferromagnetically zigzag (ZZ) ordered state. The low-field evolution of spin structure is still a matter of debate and the magnetic anisotropy within the honeycomb planes is an open and challenging question. Here, we investigate the evolution of the ZZ order parameter by second-order magneto-optical effects, the magnetic linear dichroism and magnetic linear birefringence. Our results clarify the presence and nature of metamagnetic transitions in the ZZ phase of $\alpha$-RuCl$_3$. Our experimental observations show the presence of initial magnetic domain repopulation followed by a spin-flop transition for small in-plane applied magnetic fields ($\approx$ 1.6 T) along specific crystallographic directions. In addition, using a magneto-optical approach, we detected the recently reported emergence of a field-induced intermediate phase before suppressing the ZZ order. Our results disclose the details of various angle-dependent in-plane metamagnetic transitions quantifying the bond-anisotropic interactions present in $\alpha$-RuCl$_3$.
\end{abstract}

\maketitle

\textbf{Introduction}

Quantum materials with exotic spin liquid ground state properties arose a lot of interest due to their potential in both fundamental science and application in "topological" quantum computing devices \cite{savary2016quantum, RevModPhys.80.1083}. Especially Mott-Hubbard insulating frustrated magnets with strong spin-orbit coupling and effective $j_{\text{eff}}=1/2$ states have been intensively studied as they are believed to be prime candidates realizing the physics of the exactly solvable Kitaev model on a honeycomb lattice \cite{Kitaev20062,jackeli2009mott,chaloupka2010Kitaev,trebst2017Kitaev,doi:10.1146/annurev-conmatphys-033117-053934,winter2017models}. The ground state in this exactly solvable model is a spin liquid state, i.e. a highly-entangled topological state of matter without long-range magnetic order where spin-flip excitations fractionalize into itinerant Majorana fermions and emergent gauge fields, which are believed to play a key role in fault-tolerant quantum computing \cite{Kitaev20032,nayak2008non}. Especially, the trihalide $\alpha$-RuCl$_3$ has attracted immense attention as the prime candidate to show Kitaev spin liquid physics since several experimental studies indicated fingerprints of dominant Kitaev interactions in this Mott-Hubbard insulating magnet \cite{banerjee2017neutron,baek2017evidence,kasahara2018unusual,yadav2016kitaev,banerjee2018excitations,banerjee2016proximate,do2017majorana,sandilands2015scattering,czajka2021oscillations,kasahara2018majorana,wang2020range,sears2020ferromagnetic,wen2019experimental,ran2017spin,PhysRevB.94.064435}, which was supported by multiple theoretical calculations \cite{winter2018probing,laurell2020dynamical,gordon2019theory,catuneanu2018path,suzuki2018effective,PhysRevB.96.064430}. However, despite the finite Kitaev interactions $K$, $\alpha$-RuCl$_3$ establishes long-range antiferromagnetic zigzag (ZZ) order at low temperatures, indicating the presence of significant non-Kitaev interactions, such as isotropic Heisenberg $J$ or symmetric off-diagonal interactions $\Gamma$ \cite{PhysRevB.91.144420,PhysRevB.95.180411,PhysRevB.92.235119,PhysRevB.93.134423}. 
%It is known that a magnetic field applied within the honeycomb planes can suppress the magnetic order and pushes $\alpha$-RuCl$_3$ into a quantum spin-disordered state whose properties are controversially debated \cite{PhysRevB.91.094422,PhysRevB.91.180401,PhysRevLett.119.037201}. 
To date, several experimental techniques have been applied to map out the equilibrium phase diagram of $\alpha$-RuCl$_3$ in the temperature and magnetic field plane \cite{wang2017magnetic,PhysRevB.101.140410,PhysRevB.95.180411,PhysRevB.100.060405,PhysRevB.91.094422,PhysRevB.101.245158,PhysRevB.102.214432,PhysRevLett.118.187203,PhysRevB.96.041405,lampenkelley2021fieldinduced,Balz_2021,zheng2018dielectric}. Signatures of fractionalized excitations have been detected by various spectroscopy techniques \cite{reschke2019terahertz,reschke2017electronic,banerjee2016proximate,banerjee2017neutron,banerjee2018excitations}, hinting towards a proximate spin-liquid behavior.
Initially this has lead to a wide spreading in reported values for the possible interaction strengths and a controversial discussion on the effective spin Hamiltonian capturing the experimental observations \cite{maksimov2020rethinking}. Nowadays, the parameter space of the effective spin Hamiltonian and the size and sign of the present exchange interactions converge towards a unifying description of $\alpha$-RuCl$_3$ physical properties. Especially the role of the symmetric off-diagonal exchange interaction $\Gamma$ has been studied intensively and Sears \textit{et al.} reported recently that its size is comparable to the anisotropic ferromagnetic Kitaev $K$ exchange interaction \cite{sears2020ferromagnetic}, indicating its key role in understanding $\alpha$-RuCl$_3$ large anisotropic susceptibilities for magnetic fields applied within $\chi_{\parallel}$ and perpendicular $\chi_{\perp}$ to the honeycomb planes \cite{PhysRevB.91.144420,PhysRevB.91.180401,PhysRevB.91.094422}. 
Despite this, also the orientation of a magnetic field applied within the honeycomb planes has been found to be crucial to resolve strongly angle-dependent low-energy excitations revealing fingerprints of a potential QSL state in $\alpha$-RuCl$_3$ \cite{czajka2021oscillations,tanaka2020thermodynamic,PhysRevResearch.3.013179}. Different experimental studies reported that $\alpha$-RuCl$_3$ undergoes several phase transitions at low temperatures and small finite applied fields before entering the quantum-disordered phase \cite{Balz_2021,lampenkelley2021fieldinduced,PhysRevB.98.094425,PhysRevLett.125.037202,PhysRevB.103.054440,PhysRevLett.125.097203,PhysRevB.103.174413}, and the appearence of metamagnetic transitions has been predicted theoretically \cite{PhysRevB.96.064430,janssen2019heisenberg}. The emergence of an intermediate field-induced transition at around $6.2$ T, depending on the orientation of the external in-plane magnetic field either along or perpendicular to the Ru-Ru bonds \cite{Balz_2021,lampenkelley2021fieldinduced,PhysRevLett.125.097203,PhysRevB.103.174413,PhysRevB.103.054440} has been reported and points towards the necessity to include anisotropic inter-layer exchange interactions in the model Hamiltonian. In contrast, the low-field response for in-plane fields up to $\approx 2$ T is still only partially understood, since the precise knowledge about the orientation of the order parameter is difficult to attain with thermodynamic probes. Therefore, its nature has been interpreted differently \cite{PhysRevB.98.094425,PhysRevB.91.094422,shi2018field,banerjee2018excitations,PhysRevB.95.180411}, while there are first experimental signatures revealing the necessity to consider the present bond-anisotropy stemming from a small inequivalence in the Ru-Ru bond length in $\alpha$-RuCl$_3$ \cite{PhysRevB.98.100403}. Nevertheless, the detailed magnetic-temperature ($B,T)$ phase diagram of $\alpha$-RuCl$_3$ is still under intense debate and an experimental probe that couples sensitively to the zigzag order parameter needed to be utilized.  

Magneto-optical (MO) spectroscopy is a well-established and powerful contactless technique to explore magnetic ordering phenomena and the emergence of topological magnetic structures on small scales with remarkable sensitivity \cite{versteeg2020nonequilibrium,PhysRevMaterials.5.014403,smolenskiui1975birefringence,tesavrova2012high,boschini2015,tesavrova2012direct,saidl2017optical,PhysRevB.89.085203,carpene2015,PhysRevMaterials.3.124408,versteeg2016optically,giannotti2016}. Especially the quadratic MO effects are perfectly suited to study the antiferromagnetic ordering vector ($\bf{L}=\bf{M}_\uparrow-\bf{M}_\downarrow\neq0$) and recently, direct optical probing of zigzag antiferromagnetic order via optical spectroscopy has been reported \cite{zhang2021observation}. 
The main quadratic MO effects (even in $\bf{M}$) in reflection are named magnetic linear dichroism (MLD) and magnetic linear birefringence (MLB) \cite{smolenskiui1975birefringence,eremenko2012magneto,mak2019probing}, which are defined for the reflection of linearly polarized light under normal incidence and depend on the difference in the diagonal components of the dielectric tensor \cite{PhysRevB.89.085203}. In this context, the origin of MLD and MLB can be understood in terms of different absorption (reflection) coefficients parallel and perpendicular to the magnetization $\bf{M}$ or Néel vector $\bf{L}$. These effects manifest themselves in the polarization rotation $\theta$ (MLD) of a linearly polarized light upon reflection from the sample or give rise to an elliptical polarization $\eta$ (MLB). They stem from the spin-orbit and anisotropic exchange interactions and can be related to spin-spin correlation functions \cite{smolenskiui1975birefringence,ferre1984linear,ferre1983magnetic,sano1990symmetry}. For symmetry reasons, it follows that the considered second-order MO effects are to lowest order quadratic in the antiferromagnetic order parameter such that the scaling ($\theta, \eta)\propto L^2$ holds \cite{saidl2017optical} (see SI ~\cite{SI} for more details ).
 
\begin{figure}
\includegraphics[width=\columnwidth]{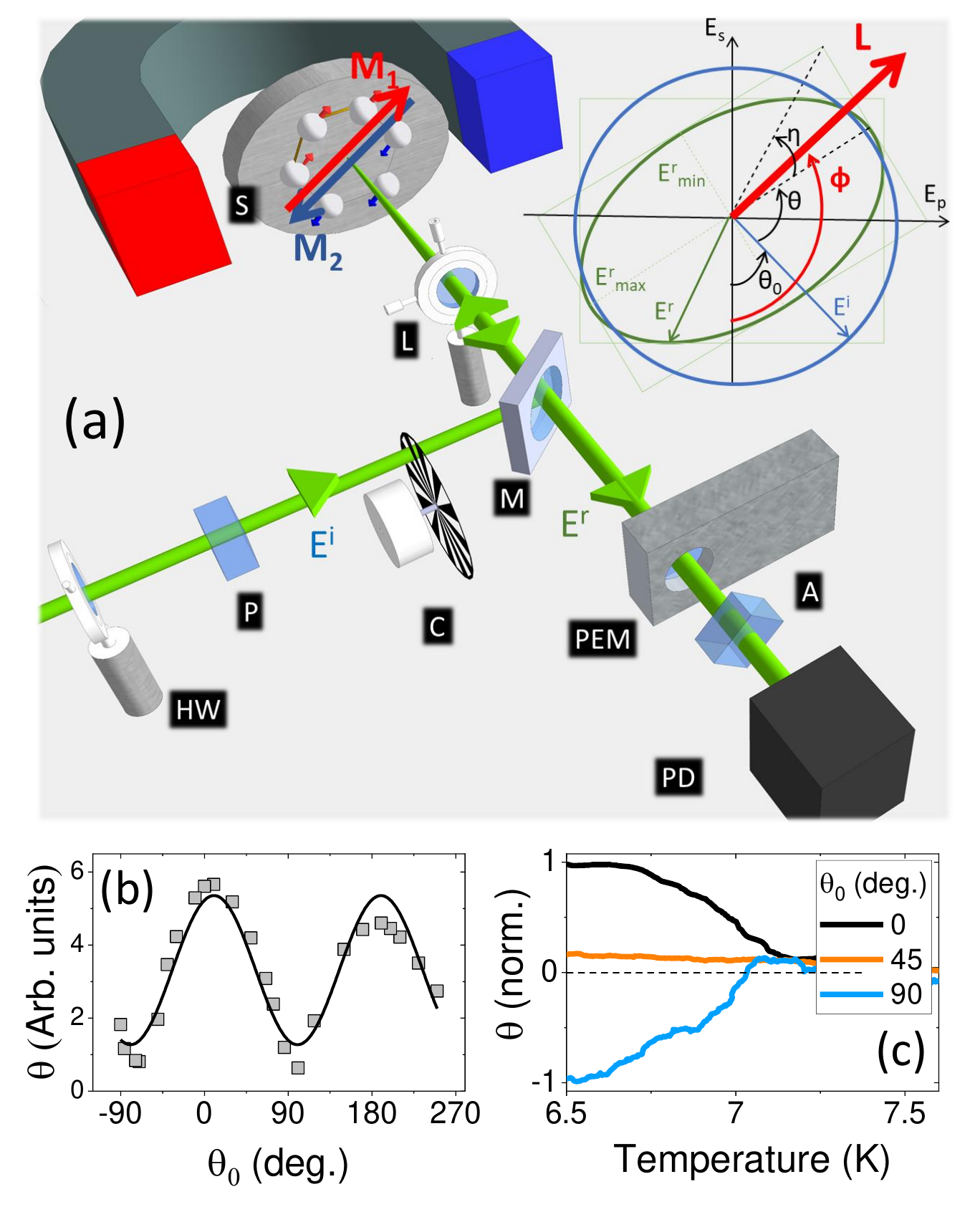}
\caption{{\bfseries{Magneto-optical experiment}}. (a) Experimental setup (HW: half-wave plate, P: polarizer, C: chopper, M: beam-splitting mirror, L: lens, S: sample, PEM: photo-elastic modulator, A: analyzer, PD: photo-diode). $E^{i}$ and $E^{r}$ correspond to the incident electric field polarization and the one upon reflection from the sample, respectively. The inset displays the effect of MO rotation $\Theta=\theta+i\eta$ and the definition of the angles $\theta_0$, $\phi$, $\theta$. (b) Polarization dependence of the MO response as a function of the polarization orientation $\theta_0$ of the incident light. Grey squares are data, solid line is a fit according to equation (\ref{eq2}). (c) Universal temperature dependence of the MO response for three different incident polarization orientations of $\theta_0=0^\circ, 45^\circ,90^\circ$.}
\label{Fig1}
\end{figure}
Motivated by these intriguing questions, we performed a systematic MO spectroscopy study to track the evolution of the ZZ order parameter in $\alpha$-RuCl$_3$ in thermodynamic equilibrium for the first time. The orientation-dependent MO response reveals the effect of the present magnetic anisotropy for magnetic fields applied perpendicular and parallel to the Ru-Ru bonds within the honeycomb layers. We show that the remarkable sensitivity of MO spectroscopy helps to clarify the emergence of two different intermediate field-induced and orientation-dependent metamagnetic transitions. Our results provide a detailed picture of the low-field behavior clarifying the influence of unidirectional bond-anisotropy within the honeycomb planes and we derive a value for the anisotropy field strength.\\

\textbf{Results}\\

First, we explore the nature of the detected MO signal and its relation to the magnetic order. Here, we apply the Voigt geometry \cite{PhysRevB.89.085203}, which is typically used to study antiferromagnets with spin alignments that are perpendicular to the light wave vector. Below, we derive the relation of the experimentally observed rotation of the polarization plane of the reflected linearly polarized light, but the same consideration applies to the change in the ellipticity. The rotation $\theta$ is related to both the amplitude of the order parameter and the relative orientation of the Néel vector to the electric field $\bf{E}$ of the incident linearly polarized light. In $\alpha$-RuCl$_3$ the in-plane component of the Néel vector is oriented parallel to the zigzag chain direction. Fig.~\ref{Fig1}(a) depicts the relative angles of the polarization of incident light and the Néel vector to the vertical polarization by $\theta_{0}$ and $\phi$, respectively. The MLD response is then given by \cite{tesavrova2012direct,tesavrova2012high}
\begin{equation}
\tan (\theta_{\text{MLD}})=\frac{\left(r_{\|}-r_{\perp}~\right) \tan \left(\phi-\theta_0~\right)}{r_{\|}+r_{\perp} \tan ^{2}\left(\phi-\theta_0~\right)}.
\label{eq1}
\end{equation}

More general and taking the presence of both linear and second-order MO effects into account, a total rotation of the polarization $\theta$, in the limit of small rotations, can be expressed as 
\begin{equation}
\theta= \mathcal{A}^{Lin}+\mathcal{A}^{MLD}\sin[2(\phi-\theta_0)].
\label{eq2}
\end{equation}
The coefficients $\mathcal{A}^{Lin}$ and $\mathcal{A}^{MLD}$ are the amplitude of linear and quadratic MO effects, respectively. Since in the MLD geometry, the incident light is normal to the surface, the $\mathcal{A}^{Lin}$ is mainly determined by the polar MO Kerr effect (PMOKE). $\mathcal{A}^{MLD}=\frac{1}{2}\left(\frac{r_\parallel}{r_\perp}-1~\right)$, where $r_\parallel$ and $r_\perp$ are the amplitude reflection coefficients of the light polarized parallel and perpendicular to the Néel vector, depends quadratically on the in-plane component of the antiferromagnetic order parameter $L$, i.e. $\mathcal{A}^{MLD}\sim L^2$ \cite{saidl2017optical} (see SI ~\cite{SI}). It is worth noting that the above relation indicates that the MLD signal, in contrast with the linear MO response, is a harmonic function of the incident polarization $\theta_0$, which becomes maximal for $\phi-\theta_0=45^\circ$ and it indicates an extreme sensitivity of the second-order MO response to the orientation of the incident linearly polarized light with respect to the spin pointing direction. Recently it has been reported that the presence of antiferromagnetic zigzag chains can give rise to a polarization dependent MO response, which is independent of the spin-pointing direction \cite{zhang2021observation}. In $\alpha$-RuCl$_3$ both, the zigzag chain direction and the spin-pointing direction are collinear in small magnetic fields, such that the spin-pointing and zigzag chain direction cause the MLD. Clearly, the MO response shown in Fig.~\ref{Fig1}(b) is polarization-dependent, which is manifested in a sinusoidal modulation as expected for MLD. At the same time, the MO response scales with $\mathcal{A}^{MLD}\propto L^2(B,T)$. Fig.~\ref{Fig1}(c) shows the temperature dependence of the zigzag order parameter studied for three different polarization orientations of the incident light, $\theta_0$. Clearly, the magnitude of $\theta$ depends on $\theta_0$ and for $\theta_0=45^\circ$ becomes zero, while the qualitative temperature dependent behavior remains, as expected, similar. In subsequent measurements we chose the $s$-polarized probe to obtain the maximum signal. 

\begin{figure*}
\includegraphics[width=0.8\textwidth]{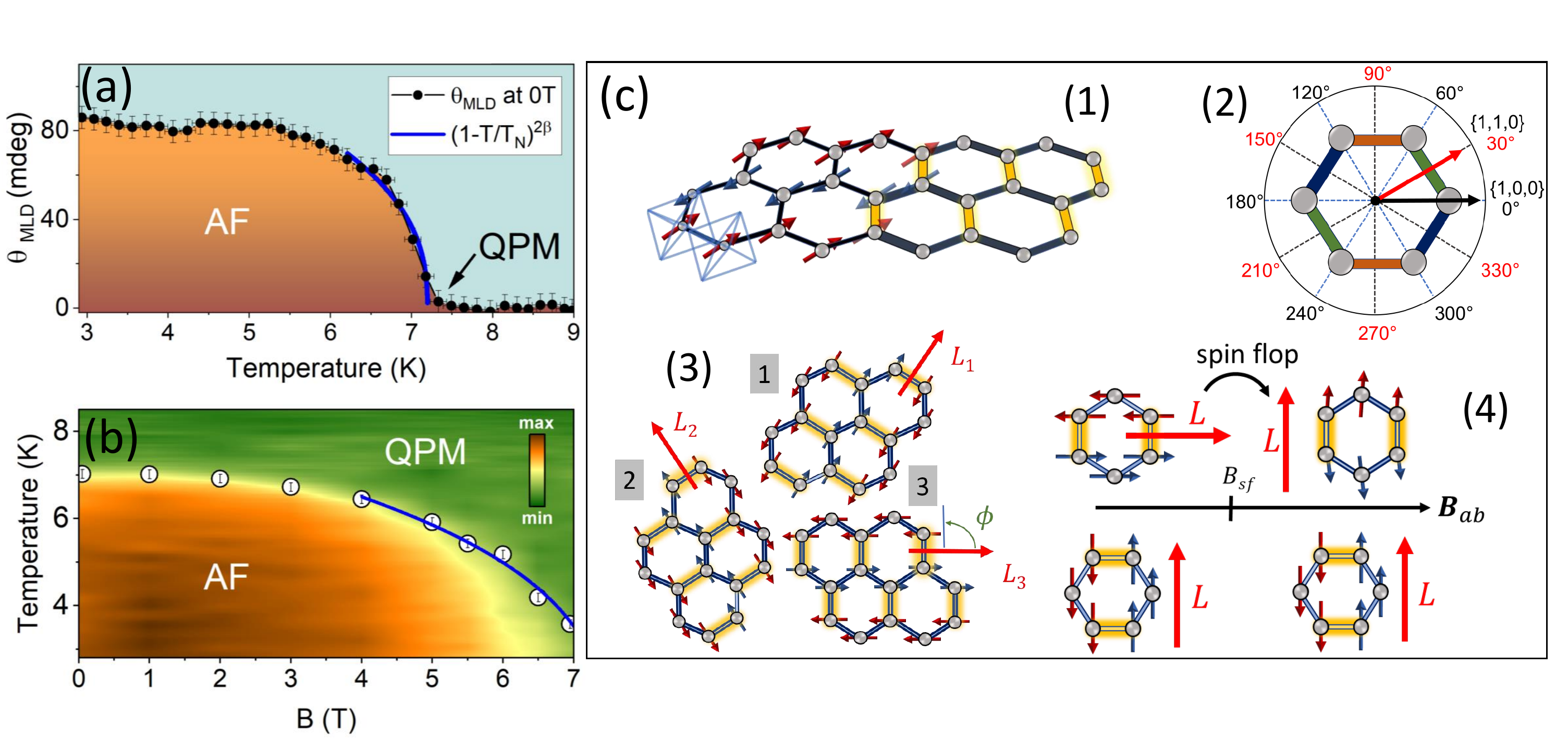}
\caption{{\bfseries{Phase diagram of $\alpha$-RuCl$_3$ and temperature-dependent magneto-optical probe of the order parameter evolution in magnetic field}}. (a) Evolution of the rotation $\theta_{\text{MLD}}$ as a function of the temperature in zero field (black dots). The blue line is a power law fit. (b) In-plane magnetic field and temperature phase diagram of $\alpha$-RuCl$_3$ constructed from the rotation $\theta_{\text{MLD}}$ obtained by iso-magnetic MLD measurements. The Néel temperature extracted from the derivative of the rotation $[(\text{d} \theta / \text{d} T)/ T]_B$ is shown as white circles. The colour code displays the amplitude of the rotation from 0 to 70 mdeg. (c) (1) Bond and out-of-plane spin orientation. The spins enclose an angle of $\sim32^{\circ}$ with the $ab$-plane. (2) Definition of the field direction along different crystallographic orientations in the honeycomb $ab$-plane. Differently coloured bonds indicate the $x,y$ and $z$-bonds of the Kitaev Hamiltonian \cite{Kitaev20062}, with interactions $J,K,\Gamma$. (3) Sketch of three possible zigzag domains projected to the 2D $ab$-plane related by a $120^\circ$ rotation of the ordering wave vector $\bf{L}$. Yellow bond is the elongated bond compared to the blue ones. (4) Sketch of the in-plane field angle dependent spin-flop transition.}
\label{Fig2}
\end{figure*}

Fig.~\ref{Fig2}(a) shows the variations of $\theta_\text{MLD}$ as a function of temperature. According to \cite{banerjee2017neutron}, the temperature dependence of the zigzag order parameter in zero magnetic field $L(T, B=0\text{T})$ follows a power-law $\theta_{\text{MLD}}\propto L^2 \propto (1-T/T_N)^{2\beta}$ below the Néel temperature $T_N$. The deduced Néel temperature $T_N=(7.19\pm0.14)$~K indicates $ABC$-stacking in the samples under study as opposed to $ABAB$-stacking accompanied by stacking faults, which causes further transitions above $T_N$ \cite{PhysRevB.103.174413}. Within the experimental uncertainty, the data can be fitted using a critical exponent $\beta=0.19\pm0.07$), which is close to the 2D Ising universality class \cite{odor2004universality}. Fig.~\ref{Fig2}(b) displays the phase diagram in the $(B,T)$ plane extracted from the MLD induced rotation at different fixed magnetic fields (iso-magnetic). We find the evolution of the order parameter $L(T, B)$ to scale proportional to $(H-H_c)^\gamma$, where $H_c$ corresponds to the critical magnetic field above which the antiferromagnetic ZZ order is suppressed and the system enters the magnetically disordered state. The power-law fit exhibits a critical field value of $H_c=(7.48\pm 0.3)$~T and $\gamma=(0.31\pm0.07)$, which is again close to the theoretical value of $0.32$ for the 2D Ising symmetry class \cite{PhysRevB.95.180411}. These findings support that the MO response clearly displays the evolution of the order parameter. More details on the fitting of the critical behavior can be found in the SI.

Fig.~\ref{Fig2}(c) displays the honeycomb structure, bond-anisotropy and spin orientation in $\alpha$-RuCl$_3$. We point out the breaking of $C_3$ symmetry in $\alpha$-RuCl$_3$ crystals originating from inequivalent Ru-Ru bond lengths, which leads to the existing monoclinic $C2/m$ space group \cite{PhysRevB.92.235119,PhysRevB.93.134423}. Phenomenologically, the inequivalence in the Ru-Ru bond lengths causes a change in the present interactions along the stretched bond indicated by $J',K'$ and $\Gamma'$ (cf. Fig.~\ref{Fig2}(c)). This causes the pseudspins, which are tilted by an angle of $\approx32^{\circ}$ out of the honeycomb plane \cite{sears2020ferromagnetic} to have their in-plane projection being preferentially oriented perpendicular to the stretched bonds. This is a key point that needs to be taken into account to understand the anisotropic MO response in the following results. Nevertheless, the local $C2/m$ symmetry has been found to be broken in multi-domain samples \cite{PhysRevB.98.094425}, due to a randomness in the monoclinic distortion of one Ru-Ru bond. Consequently, there can be three possible and symmetry-allowed zigzgag domains in zero field cooled samples below $T_N$, which are related by $120^\circ$ rotations within the plane. However, applying a finite magnetic field along specific crystallographic orientations can change the spin pointing direction within the honeycomb planes as will be discussed in the following and is schematically depicted in Fig.~\ref{Fig2}(c) via a spin-flop process.

Having established the iso-magnetic response of the order parameter $L$ encoded in the MLD response $\theta_\text{{MLD}}$ we turn now to the iso-thermal MLD response of $\alpha$-RuCl$_3$ at a temperature of $3$~K and magnetic field strengths up to $\pm7$~T applied within the honeycomb planes along two different crystallographic directions to investigate the magnetic in-plane anisotropy. We studied two samples from the same batch which have been oriented along different crystallographic directions. Once, for sample (a) the in-plane magnetic $\mathbf{B}_{ab}$ field is applied perpendicular to a Ru-Ru bond, i.e. along one of the symmetry-equivalent $\{1,1,0\}$ directions, while the field was directed parallel to the Ru-Ru bonds for sample (b), i.e. along one of the symmetry-equivalent $\{1,0,0\}$ directions (see Fig.~\ref{Fig2} (c)). The MO measurements were conducted with the magnetic field swept continuously from $0$ to $\pm7$ T to systematically track the dynamical MO response (see SI \cite{SI} for more information).

\begin{figure*}
\includegraphics[width=0.85\textwidth]{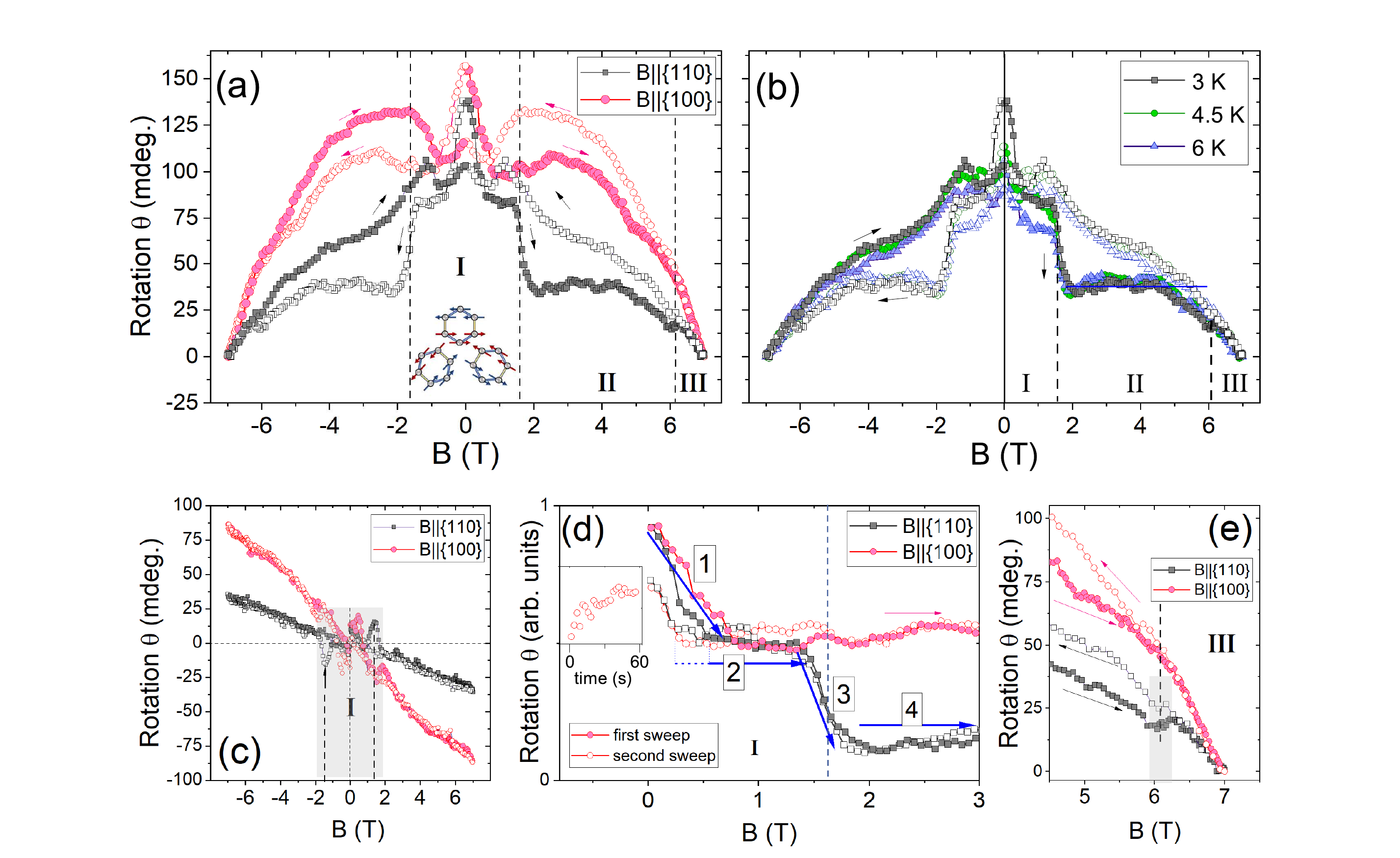}
\caption{{\bfseries{Magnetic linear dichroism of $\alpha$-RuCl$_3$} for different sample orientations.} (a) Full magnetic field sweep scans of the magneto-optical response for $\bf{B_{ab}} \parallel{\{1,1,0}\}$ and  $\bf{B_{ab}}\parallel{\{1,0,0\}}$ at 3 K. The hysteresis loops can be divided into three field regimes \Romannum{1}, \Romannum{2} and \Romannum{3}, which are discussed in the main text. (b) Temperature dependence of the MLD response as the magnetic field is aligned along the ${\{1,1,0\}}$ direction. Labels are discussed in the main text. (c) Linear MO response extracted from the MLD response for the two different field orientations. The light grey area indicates field-regime \Romannum{1}. (d) Zoom-in into field regime \Romannum{1} and \Romannum{2} for two subsequent field sweeps, which are divided into 4 steps as discussed in the main text. Inset shows the zero field  time-dependent recovery of MO signal after the field sweep. (e) High field regime \Romannum{3}. The kink in the rotation for $\bf{B_{ab}}\parallel{\{1,1,0}\}$ highlighted by the dashed line is connected to the transition towards the new AFM phase. The shaded area indicates the emergence of the intermediate magnetic transition.}
\label{Fig3}
\end{figure*}

Fig.~\ref{Fig3}(a) shows that the MO response $\theta_\text{MLD}$ differs significantly for two distinct orientations of an in-plane magnetic field $\bf{B_{ab}}$, a first experimental evidence for the present in-plane magnetic anisotropy.
The purely magnetic origin for this anisotropy is verified by the fact that the temperature-dependent evolution of $\theta_{\text{MLD}}$ and $\eta$ in zero magnetic field is similar for both samples ruling out possible temperature-related effects like strain or thermoelastic changes (see SI \cite{SI}). The absolute field-induced change in the rotation defined as $\Delta \theta(B)=\theta_{\text{MLD}}(0\text{T})-\theta_{\text{MLD}}(\pm7\text{T})$ for both field configurations at $3$ K is in the order of $100$ mdeg, which is large and underlines the microscopic impact of strong spin-orbit interactions ($\sim 100$ meV in $\alpha$-RuCl$_3$ \cite{PhysRevB.93.075144}) on the second order MO responses. The value at 7 T was set to zero to extract the field-induced changes in the rotation between 0 and 7 T at constant temperature (see SI). Despite the difference in the full hysteresis loops for the different field orientations, we divide the MO response shown in Fig. \ref{Fig3}(a) into three field regimes to allow a simple description and comparison. In each regime, the MO signal is comparable, while at the crossovers a vivid anisotropic response is present. The first transition at $\pm1.6$ ~T is a large and steep change in the rotation $\theta_{\text{MLD}}$ (absolute reduction of $\approx$ 50 mdeg) for the magnetic field applied perpendicular to the Ru-Ru bonds, whereas the rotation changes only slightly for the in-plane magnetic field applied parallel to the bonds. The marked difference for the two differently oriented samples in regimes \Romannum{1} and \Romannum{2} is displayed in Fig.~\ref{Fig3}(d) in more detail. This clear difference indicates the presence of a metamagnetic transition as will be discussed later. The second transition in the response occurs at around $6.2$~T for the field applied perpendicular to the Ru-Ru bonds where an anomaly in $\theta_{\text{MLD}}$ is observed by a kink. Reducing the magnetic field strength continuously from $\pm7$~T back towards $0$~T causes the emergence of significant hysteresis in the MLD response opening at a field strength of $\approx 6$~T. The hysteresis loops close at a field strength of $\approx 0.7$~T for both field orientations, but open again for small fields close to $0$~T. Here, the integrated hysteresis weight is a factor of $\approx 1.5$ larger for the field applied perpendicular to the bonds than along the bonds pointing again towards a difference in the in-plane anisotropy energy.\\
Fig.~\ref{Fig3}(b) reports the iso-thermal MO responses as a function of the applied magnetic field. We found a similar behaviour showing the three regimes \Romannum{1}-\Romannum{3} for three bath temperatures of $3$~K, $4.5$~K, and $6$~K, corresponding to locations in the phase diagram deep insight the zigzag phase, in an intermediate range and close to the critical transition temperature towards the quantum paramagnetic phase. This finding indicates the purely magnetic nature of the MO field response.\\
In contrast, the odd MO response scales linearly in the applied magnetic field for both samples and shows no pronounced hysteresis for a whole field sweep (Fig.~\ref{Fig3}(c)). However, in regime \Romannum{1} small spikes can be observed for both samples. 

\begin{figure*}
\includegraphics[width=0.83\textwidth]{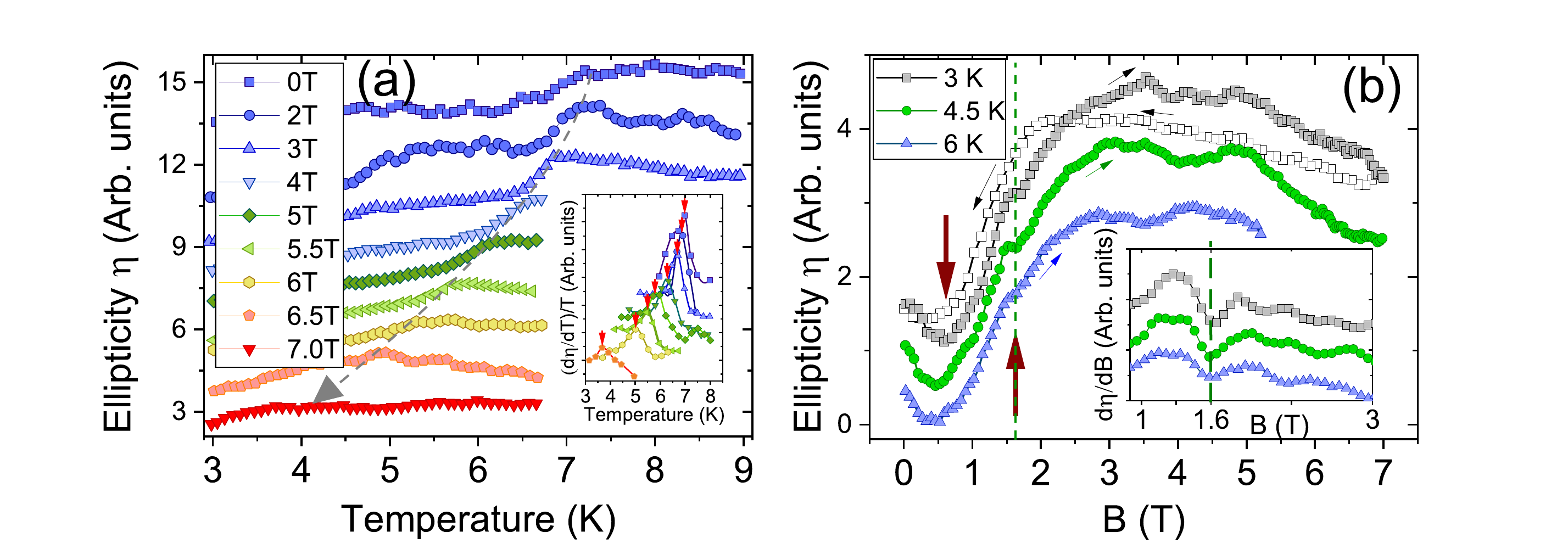}
\caption{{\bfseries{Magnetic linear birefringence of RuCl$_3$.}} (a) Iso-magnetic MLB as a function of the temperature. Data is plotted offset for clarity. The inset shows the calculated first derivative of ellipticity $(\text{d} \eta / \text{d} T)$ normalized to the bath temperature. Red arrows indicate the maximum of the derivative indicating the magnetic phase transition, which is indicated by the grey dashed arrow in the main figure. %(b) The extracted Néel temperatures from $[(\text{d} \eta / \text{d} T)/ T]_B$ and the power law fit in blue. 
(b) MLB $\eta$ for $\bf{B_{ab}}\parallel{\{1,1,0}\}$ as a function of the in-plane magnetic field at different temperatures. The arrows indicate the anomalies discussed in the main text. %Arrows indicate changes in $\eta$, which can also be observed in the standard deviation of $\text{d}M(B) $ in Fig.~\ref{Fig5}(b). %(d) MLD response of sample (a) as a function of in-plane magnetic field for different temperatures. The numbers indicate the steps in regime \Romannum{1}. (e) Magnetization measurement for both samples. The field was oriented in the $ab$-plane. The inset shows $\text{d} M / \text{d} B$ and arrows indicate anomalies discussed in the main text. (f) Standard deviation of the magnetization measurements shown in (e). Details are discussed in the main text.
}
\label{Fig4}
\end{figure*}
In order to understand the nature of this spikes, MOKE measurements with the sample rotated by $45^\circ$ w.r.t. the incident light wave vector have been performed. We find a clear featureless linear scaling of the obtained rotation with the applied magnetic field (see SI \cite{SI}). Therefore, we assume that the small spikes are related to second-order MO effects. Since under the real experimental condition, the perfect symmetry cannot be achieved, minor differences in MLD obtained from negative and positive fields are unavoidable.\\ Fig.~\ref{Fig3}(d) compares the initial part of the hysteresis in regime \Romannum{1} for both crystallographic orientations for subsequent field sweeps from 0 to 7 T. To emphasize similarities, we consider the normalized MO response. Here, we can divide the field dependent alternation of MLD into he following four distinct steps.\\
(i) The first step shows the following characteristics. First, it displays an initial big, but gradual and monotonic change in $\theta_\text{MLD}$, which starts immediately when applying a small in-plane magnetic field. Second, it is seemingly independent on the field orientation in the honeycomb plane and terminates at a field strength of $\sim 0.7$~T. Third, the absolute amplitude of this first step decreases for subsequent field sweeps, but displays a similar field dependence for both field orientations. Fourth, the MO response does not recover immediately when sweeping the field back to 0 T. However, there is a slow recovery of the MO response within tens of seconds at zero field to a saturating value (see inset \ref{Fig3}(d)). The above mentioned distinct features of step 1, as we explain later, can reasonably be assigned to an initial field-driven gradual domain repopulation or zigzag domain switching.\\
(ii) In step 2, no further change in the magneto-optical response can be observed for any of the subsequent field scans between $0.7-1.6$~T with distinct in-plane field orientations. This indicates that in step 2 neither the zigzag chain orientation nor the spin-pointing direction and the amplitude of $L$ are changed by the external magnetic field. This observation shows that at around $0.7$ to $1.6$~T the magnetic field changes between the three degenerate domains are terminated.\\
(iii) A further increase in the {applied} field strength leads then to a sharp change (step 3) at $\approx1.6$~T for the in-plane field aligned perpendicular to the bond. In contrast, this abrupt change is not present for the field directed along the bond. This is a clear indication for a field-induced metamagnetic transition, which we will discuss later in terms of a spin-flop transition originating from the intra-layer bond-anisotropy of the frustrated honeycomb magnet. This anisotropic change in rotation at a field of $\approx 1.6$~T is, in contrast to step 1, completely reproducible in subsequent field sweeps. Further, step 3 is clearly observed and remains sharp at all temperatures demonstrating the field evolution of the MO response is reproducible in the entire ZZ phase (cf. Fig.~\ref{Fig3}(b)).\\
(iv) Further strengthen the applied field leads to the emergence of another steady state of the obtained signal from 2 to 3 T (step 4) pointing towards a homogeneously field-aligned ZZ ordering. Here, similar to step 2, neither the zigzag chain nor the spin pointing direction change. Moreover, the different steps can be clearly observed in subsequent measurements at different temperatures as shown in Fig.~\ref{Fig3}(b).

These different observations confirm the magnetic origin of the field response and provide first important key information to distinguish both the effects of field- and anisotropy-related domain selection and an accompanied spin-flopped phase for in-plane applied magnetic fields. In the first step, the in-plane field leads to an immediate population of zigzag domains with the easy zigzag chain axis closer to the field direction \cite{PhysRevB.96.214418,lampenkelley2021fieldinduced,Balz_2021,PhysRevB.98.094425,sano1990symmetry}. This interpretation is further supported by a recent study of the thermal and magneto-elastic properties of $\alpha$-RuCl$_3$ for field applied in the $ab$-plane. It was found, that under an in-plane magnetic field $\alpha$-RuCl$_3$ shows lattice contraction along the $\{1,1,0\}$ direction \cite{PhysRevB.102.214432}. Especially in low fields up to $\approx 0.5$ T the magnetostriction coefficient changes continuously, such that the initial zigzag domain structure and distribution will be gradually changed by the external magnetic field. The fact that the MO hysteresis does not close while reducing the in-plane field back to $0$~T points towards irreversible processes. This fits to the initial domain repopulation picture and hysteric magnetocaloric measurements below $T_N$ \cite{PhysRevB.102.214432}.

Turning to the high-field regime \Romannum{3}, the kink which is observed only for one direction (see Fig.~\ref{Fig3}(e)) indicates clear dependence on the orientation of $\bf{B}_{ab}$ with respect to the pseudospin-bonds. This kink in the rotation is reproducible during different measurement cycles and likely related to the previously reported first-order transition into an intermediate differently ordered ZZ phase for in-plane magnetic fields of $\approx 6-6.5$~T aligned perpendicular to the Ru-Ru bond \cite{lampenkelley2021fieldinduced,Balz_2021}. We observed this kink only at low temperatures since its intensity dramatically decreases by temperature (see Fig.~\ref{Fig6}(c)) in agreement with the recent reports \cite{schonemann2020,Balz_2021}. We point out that signatures of this metamagnetic transition seem much weaker compared to the spin-flop transition at a field of $\approx1.6$~T, which is related to the already suppressed amplitude of $L$ close to the critical line and the fact that it might originate from a competition of anisotropic inter-layer exchange interactions as opposed to a change in the in-plane magnetic ordering \cite{Balz_2021}.\\

In the following, we present the results of two individual experiments which further support the observed anisotropic MLD response at $\approx1.6$~T. Fig.~\ref{Fig4}(a) shows the temperature dependence of the iso-magnetic MLB $\eta$. Since the birefringence originates from spin-spin-correlations contributing to the magnetic energy, its derivative scales proportional to the magnetic part of the specific heat \cite{PhysRevB.37.5483,PhysRevB.34.479,smolenskiui1975birefringence}. The obtained curves for $[(\text{d} \eta / \text{d} T)/ T]_B$ are shown in the inset of Fig.~\ref{Fig4}(a), which resolve changes in $\eta$ more clearly and display a typical $\lambda$-shape as has been reported previously for the magnetic part of the specific heat at the transition for $\alpha$-RuCl$_3$ \cite{PhysRevB.99.094415,kasahara2018majorana}. The red arrows indicate the peak values of these derivatives, which coincide with the vanishing MO rotation reported in Fig.~\ref{Fig2}(b). As expected, the power-law fit $(H-H_c)^{\gamma_{\eta}}$ gives similar values $H_c= (7.21\pm 0.09)$~T and $\gamma_{\eta}=(0.29\pm0.04)$ as for the MLD response. Indications of the transition at $1.6$~T are also visible in the temperature-dependence of $\eta$, where a small kink in the curves is indicated by a dashed line in Fig.~\ref{Fig4}(b) for all three temperatures.

\begin{figure}
\includegraphics[width=\columnwidth]{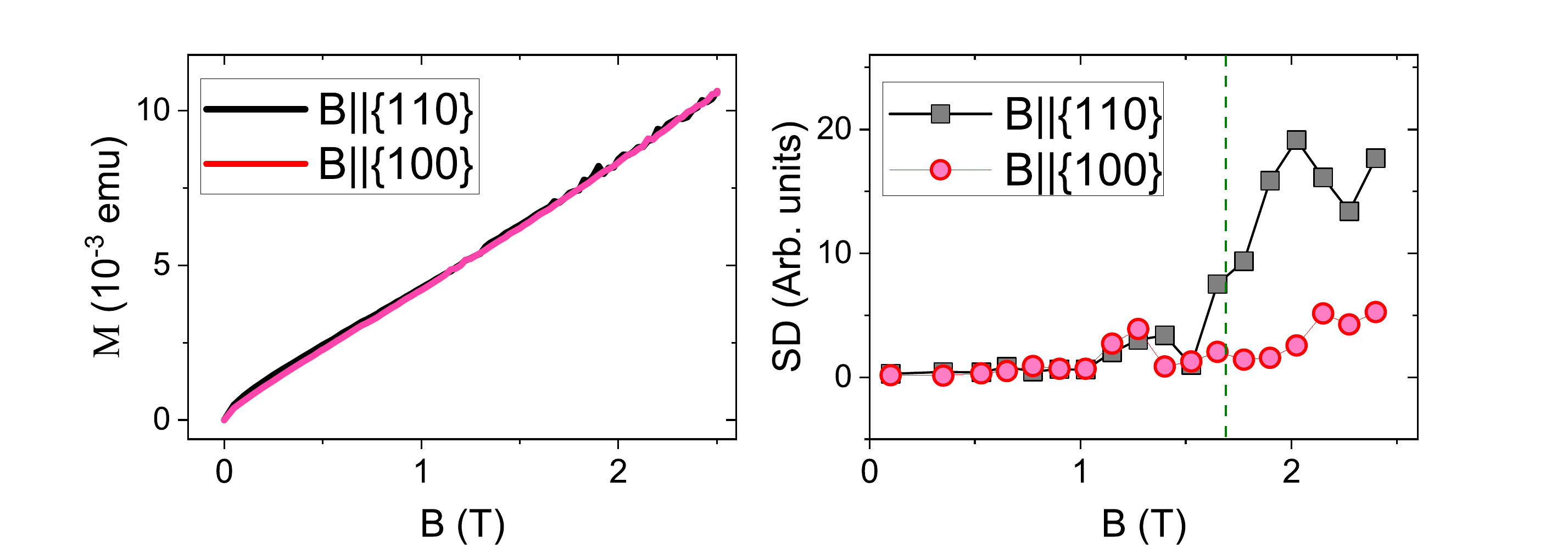}
\caption{(a) Magnetization measurement for both field orientations within the $ab$-plane. (b) Standard deviation of the magnetization measurements. Details are discussed in the main text.}
\label{Fig5}
\end{figure}

\begin{figure*}
\includegraphics[width=0.74\textwidth]{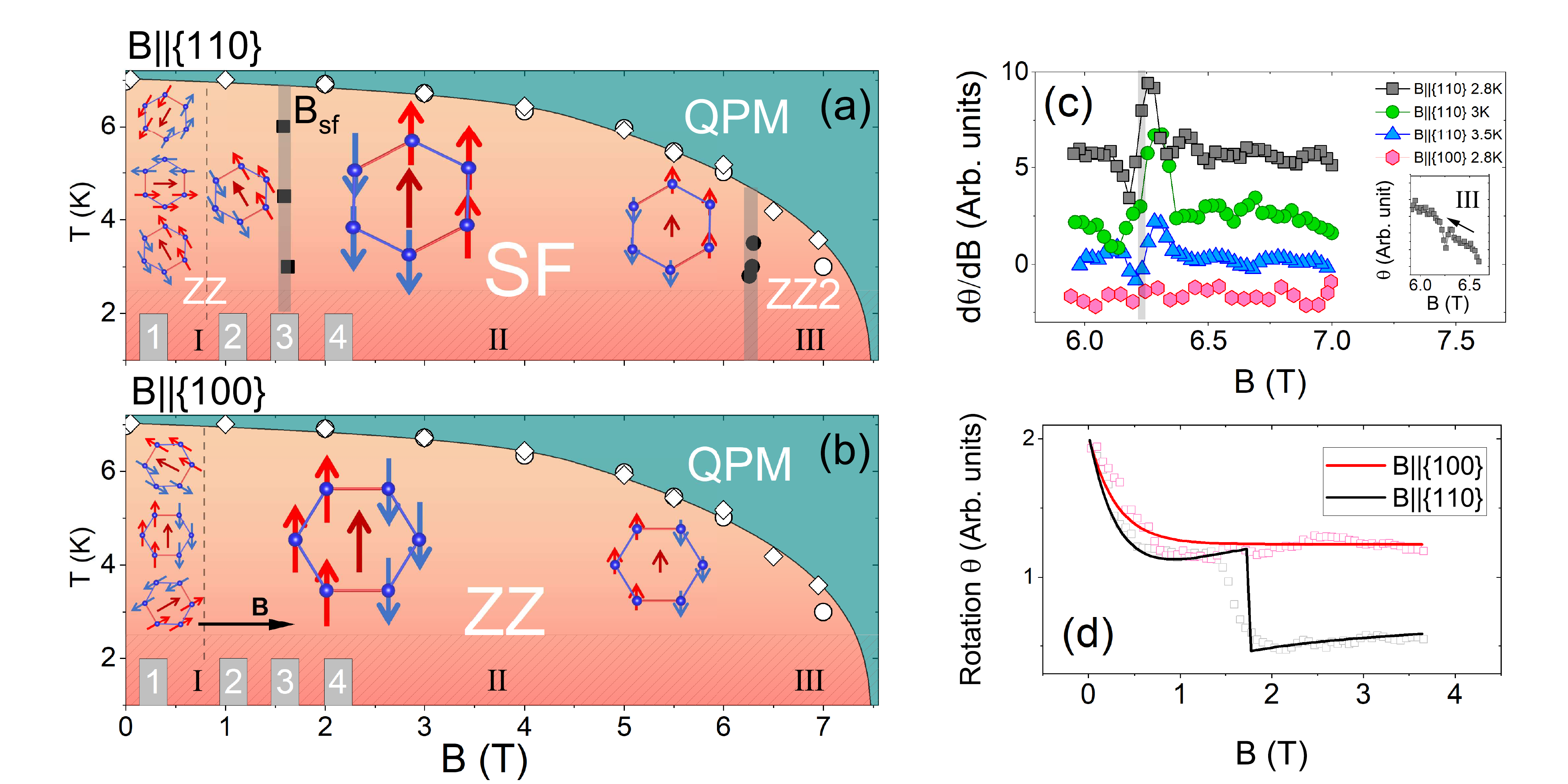}
\caption{{\bfseries{Low-field and high-field anomalies in the magneto-optical response of $\alpha$-RuCl$_3$}.} (a) and (b) Schematic sketch of the temperature-field phase diagram of $\alpha$-RuCl$_3$ for two directions, ${\{1,1,0\}}$ (B perpendicular to the Ru-Ru bond) and ${\{1,0,0\}}$ (B along the bond). The three different observed regimes \Romannum{1} (ZZ), \Romannum{2} (SF) and \Romannum{3} (ZZ2) are displayed. In regime \Romannum{1}, three initial domains are present. More details on each step (1 to 4) can be found in the main text. After step 2, the most populated domain is depicted for each field direction. For the fields larger than B$_{SF}$, the similar spin orientation is shown. Black square and circle symbols are derived from MLD experiments and show the metamagnetic transition fields for different temperatures. White square and circle symbols show the boundary with QPM phase obtained by MLD and MLB experiments, respectively.
(c) High field anomalies of  $\text{d} \theta / \text{d} B$ at different temperatures. The inset shows the jump in the MLD data at $6.2$ T. (d) The proposed model simulates domain population and the spin-flop transition.}
\label{Fig6}
\end{figure*}

Furthermore, we performed in-plane magnetometry measurements at $3$~K (see Fig.~\ref{Fig5}(a)) using a superconducting quantum interference device (SQUID, Quantum Design, MPMS). The obtained magnetization curves for both field orientations as in the MO experiments cross through the zero point, as expected for a compensated antiferromagnet, excluding any finite ferromagnetic contribution or background signal. Although at first glance, the magnetization curves $M(H)$ do not exhibit any clear signature of a transition at $1.6$~T, the standard deviation of the magnetization measurements illustrated in Fig.~\ref{Fig5}(b) give an indication (for more details see SI \cite{SI}). The spin-flop transition is of first-order, i.e. close to the critical field value both the initial zigzag-oriented domains and some already spin-flopped domains coexist, such that the system is driven into a regime of large fluctuations stemming from the competition of anisotropy-stabilized zigzag and already field-driven spin-flopped domains. These fluctuations display the instability of the coexistence of energetically different zigzag phases near the critical spin-flop field strength, which leads to discontinuous jumps in the magnetization accompanied by irreversible behavior. The instability is clearly visible in the standard deviation of the magnetization measurements illustrated in Fig.~\ref{Fig5}(b). This effect has been observed previously in susceptibility measurements at the phase boundary between the antiferromagnetic and spin-flopped phase in the hexagonal antiferromagnet NiO accompanied by an initial field-induced domain alignment \cite{machado2017spin}.\\

\textbf{Discussion}\\

Based on these independent and consistent experimental observations, we elaborate on our main findings of the response to the magnetic field orientation along the two different crystallographic orientations and discuss the low-field response in detail. In zero field, the sample comprises three possible zigzag domains in which the spin direction differs by $120^\circ$ with statistically distributed unequal but likely comparable populations. A small external applied field within the $ab$-plane will then favor zigzag domain(s) for which the Néel vector is most nearly perpendicular to $\bf{B}_{ab}$ in order to maximize the susceptibility. Such a scenario has been discussed for a similar three-fold degenerate domain structure in the uniaxial antiferromagnet NiO \cite{machado2017spin}. This is perfectly in line with the initial changes in the MO response indicative of the metastability of the domain population. 

During the process of field-induced domain repopulation the volume of the three distinct zigzag chains changes, which immediately is reflected in changes of $\theta_{\text{MLD}}$ \cite{zhang2021observation}. However, if this domain repopulation would be the only mechanism causing the zigzag degeneracy to be lifted, a further increase of the magnetic field strength should then only affect the amplitude of the order parameter $L$, while its orientation should be unchanged for small enough field strengths. It follows that no additional abrupt and anisotropic changes in $\theta_{\text{MLD}}$ would be expected for just a gradual repopulation of the ZZ domains. In this regard, the key information provided by the MLD response is encoded in steps 2 and 3. In step 2 the system is approaching a steady-state in an interval of the applied field from $\approx 0.7$ to $1.6$~T, that is the initial and field-orientation independet domain reorientation is terminated. Hence, the clear difference at $1.6$~T for the different orientations of the magnetic field gives experimental evidence that it has to be caused by a different mechanism than the already terminated domain repopulation. At $1.6$~T an abrupt change for the field perpendicular to the bond appears, whereas no clear change is observable for the field along the bond. 
Hence, there is seemingly a competition between the external applied field and the anisotropic intra-layer interactions stabilizing the orientation of the Néel vector in some domains untill the external field overcomes the anisotropy energy. This competition can be modelled based on a modified spin-flop theory and allows to derive a value for the anisotropy constant $K_a$ (for more details see section V in the SI \cite{SI}). We start with the free energy of the system at the ground state, which gives information about the spin pointing direction and can be generally expressed as
\cite{sano1990symmetry,sano1989basal,zhang2015spin}

\begin{equation}
    F=-\frac{1}{2}\mu_0\left(\chi_+\sin^2(\alpha)+\chi_-\cos^2(\alpha)\right)H^2+K_a\sin^2(\alpha-\Psi).
    \label{modell}
\end{equation}

Here, $\mu_0$ is the vacuum permeability, $\chi_+$ and $\chi_-$ correspond to the extrema of the in-plane oscillating susceptibility $\chi_\parallel(\phi)$ of $\alpha$-RuCl$_3$, where $\chi_+$ and $\chi_-$ occur for the magnetic field applied parallel or perpendicular to one of the Ru-Ru bond directions, respectively \cite{PhysRevB.98.100403}. The values of $\chi_+$ and $\chi_-$ at a temperature of $2$ K have been reported in \cite{PhysRevB.98.100403}. The angle of the pseudospins and the magnetic field is parametrized by $\alpha$ and $\Psi$ is the angle between the applied field and the magnetic easy axis, which is perpendicular to one of the stretched Ru-Ru bonds. Equation (\ref{modell}) was numerically evaluated for each single zigzag domain and field orientation to find the angle $\alpha$ at fixed $\Psi$, which minimizes the energy for each field value $H$. For this, $\alpha$ was varied between $0$ and $\pi$ for each field value $H$. For the field oriented along the stretched bond, $\alpha=\pi/2$ does not change for increasing field. For anisotropy-stabilized domains at a finite angle with the applied field we find a continuous change in $\alpha(H)$, which converges towards $\alpha=\pi/2$ for increasing field strength. Only for the field applied perpendicular to the stretched bond there is a discontinuity in $\alpha(H)$. In addition, we consider the initial field-induced changes in the zigzag domain volumes on a phenomenological level similar as reported in \cite{sano1990symmetry}. More information regarding the modeling of the MO response can be found in the SI \cite{SI}.

The model satisfactorily reproduces the field evolution of the MO response for both in-plane field directions (see Fig.~\ref{Fig6}(d)). Further, the anisotropy constant $K_a$ can be estimated from the experimentally determined spin-flop field $H_\text{sf}=1.6$ T applied within the $ab$-plane
\begin{equation}
    K_a=\frac{1}{2}\mu_0(\chi_+-\chi_-) H^2_\text{sf}.
\end{equation}
The derived value for the effective anisotropy field $K_a/ \mu_{0}$ is $0.0027$ T. The interpretation of a spin-flop transition is supported by recently reported in-plane susceptibility measurements, where a two-fold oscillation has been observed for fields below $2$~T pointing towards the intrinsic in-plane bond-anisotropy with a $\pi$-periodicity as opposed to a characteristic six-fold oscillation emerging for larger field strengths indicating the field-induced reorientation of the order parameter and hence spin-pointing direction in $\alpha$-RuCl$_3$ \cite{Balz_2021,lampenkelley2021fieldinduced,PhysRevB.98.100403}. However, this produces only little changes in the magnetization (cf. Fig.~\ref{Fig5}(a)), although the orientation of $L$ changes remarkably. This is a reasonable explanation why in previously reported magnetization data, which lack the sensitivity to small changes in $L$, a change in the low-field susceptibility has been overlooked, although some changes at around 1.5 T in $\text{d} M / \text{d} B$ have been found previously (SI of \cite{PhysRevB.96.041405}, \cite{PhysRevB.91.094422}). \\
Fig.~\ref{Fig6}(a) and (b) summarize the experimental results. Independent of the in-plane field orientation there is an initial change in the ZZ domain population in regime \Romannum{1}. According to the direction of the applied field, some of the initial domains become more populated at the expense of the others, but the general evolution of the rotation $\theta_{\text{MLD}}$ in small fields is similar for both field orientations. The experimental data shows that this process terminates at a field strength of approx. 0.7 T. For the field range of $0.7-1.6$~T, i.e., step 2, there are already energetically optimal field-aligned domains, although some anisotropy stabilized domains exist for the field applied along the $\{1,1,0\}$ direction. Then at the transition into regime \Romannum{2} at step 3, the external field wins the competition over the bond anisotropy for the magnetic field applied perpendicular to the Ru-Ru bonds, such that the pseudospins in all domains rotate perpendicular to the field directions pushing $\alpha$-RuCl$_3$ into a single polarized spin-flopped state. A further increase in the field strength, i.e. moving closer to the critical line of the phase transition towards the quantum paramagnetic state, results in a natural decrease of the observed rotation, i.e. the magnitude of the order parameter. Before entering the quantum paramagnetic state above $7$~T another anomaly is seen in the MO data. In contrast to the low-field transition, this anomaly at $6.2$~T in $\text{d} \theta / \text{d} B$ is not immune to temperature changes and fades off for increasing temperatures from $2.8$~K to $3.5$~K systematically for fields applied perpendicular to the bond (c.f. Fig~\ref{Fig6}(c)). Its dependence on the in-plane field angle is illustrated by the fact, that for $\bf{B}_{ab}\parallel\{1,0,0\}$ no signature of an intermediate field-induced transition in the vicinity of $6.2$~T can be identified since $\text{d} \theta / \text{d} B$ is almost constant even at the lowest temperature of $2.8$~K. The fact that this intermediate field-induced transition strictly depends on the orientation of $\bf{B}_{ab}$ w.r.t. to the crytallographic axis conveys the emergence of an anisotropy related novel magnetically ordered phase. This phase was dubt ZZ2 and its origin has been discussed in terms of inter-layer anisotropies very recently \cite{Balz_2021}. It was understood in terms of a field-induced inhomogeneous spin canting between different 3-fold and 6-fold zigzag stackings for $\bf{B}_{ab}\parallel\{1,1,0\}$, while the canting is homogeneous for $\bf{B}_{ab}\parallel\{1,0,0\}$ \cite{Balz_2021}. In addition, it has been discussed in terms of an \textit{inverse} melting phenomenon \cite{PhysRevB.103.054440}. It is worth to note that the MO response is sensitive to this transition and its first-order character fits to the observed large MO hysteresis effect. Nevertheless, a microscopic origin of this field-induced transition, nor an estimate for the underlying inter-layer exchange interactions can be given based on our observations, but calls for future MO measurements at elevated magnetic fields.\\

\textbf{Conclusion}\\

In summary, we have reported the first magneto-optical spectroscopy measurement on the Kitaev spin liquid candidate material $\alpha$-RuCl$_3$. Our study establishes magneto-optical spectroscopy as a versatile experimental tool to elucidate exotic phases of quantum materials. We observed two intermediate metamagnetic transitions at $\approx1.6$~T and $6.2$~T for the magnetic field applied perpendicular to the Ru-Ru bond, while none of these transitions appears for the magnetic field aligned parallel to the Ru-Ru bonds. We clarified the nature of the low-field transition and discussed it in terms of a spin-flop transition, where the external field overcomes the anisotropy energy to align the Néel vector nearly orthogonal to the field direction. The effective anisotropy field has been determined to be $0.0027$ T. Further, we confirmed the emergence of the previously observed  high-field intermediate phase. Our results point out the importance of anisotropic intra- and inter-layer bond anisotropies and the necessity to include those in future theoretical calculations. Besides that, we illuminated the importance of the in-plane field angle, which calls for future MO studies in the high-field proximate spin liquid phase of $\alpha$-RuCl$_3$. The spin-flop transition at a moderate field strength motivates further studies on $\alpha$-RuCl$_3$, as the reorientation of the Néel vector opens pathways to vary the magnetoresistance almost continuously. Here, measurements of the anisotropic magnetoresistance, which are also even in the magnetization as MLD and MLB \cite{tesavrova2012high}, should be considered for future experiments to access the precise control of the Néel vector \cite{kriegner2016multiple}. This could in principle lead to future implementation of $\alpha$-RuCl$_3$ in antiferromagnetic spintronics \cite{jungwirth2016antiferromagnetic}.\\

\textbf{Methods}\\

{\bfseries{Sample growth, characterization and orientation}}. High-quality $\alpha$-RuCl$_3$ crystals were prepared by vacuum sublimation \cite{PhysRevB.101.140410}. The different samples of the same batch were characterized by SQUID magnetometry, showing a sharp transition at around $T_N\approx 7$ K in zero applied field corresponding to the $ABC$ stacking order \cite{PhysRevB.103.174413,PhysRevB.93.134423}. No additional magnetic transitions above $T_N$ are observed, which have been related to a different stacking order $ABAB$ with a two-layer periodicity or strain-introduced stacking faults due to extensive handling or deformation of the crystals. This bulk technique can only provide the first indication of sample quality for an optics study. Cleaving or polishing introduces strain. In this regard, we refrained from any sample treatment and used an as-grown $\alpha$-RuCl$_3$ sample. The temperature dependence of the equilibrium MO response shown in Fig.~\ref{Fig1}(d) shows a clear phase transition at $T_N\approx 7$ K, confirming good sample quality.\\
The different $\alpha$-RuCl$_3$ samples have been oriented via a standard x-ray Laue-diffractometer at room temperature.\\

{\bf{Experimental procedures and measurement technique}}.
For the study of the MO response of $\alpha$-RuCl$_3$, the high-quality as-grown samples were placed in a helium-cooled cryostat (Oxford Spectromag) with temperatures down to 2.2 K inside the coils of a superconducting magnet with magnetic field strengths up to $\pm7$~T. Fig.~\ref{Fig1}(a) illustrates the experimental setup. The magnetic field was applied along different crystallographic directions within the crystallographic $ab$-plane, i.e. within the honeycomb layers. The polarization of incident light was rotated by a half waveplate and after initial tests, set to the purely linearly $s$-polarized setting, i.e. $\theta_0=0$ for the maximum signal in zero field. The measurements of the second-order MO response were carried out in the so-called Voigt geometry \cite{PhysRevB.89.085203} at near-normal incidence, such that the light wave propagation $\bf{k}$ was perpendicular to the honeycomb layer planes ($\bf{k}\perp ab$) and magnetic field vector $\bf{B}_{ab}$ (see Fig.~\ref{Fig1}(a). The MO measurements were performed using a continuous laser with a wavelength of 532 nm and the laser spot was focused to a spot size of 200 $\times$ 200 microns on the sample with the power set to \SI{50}{\micro\watt}. Detection of the MO response $\Theta=\theta+i\eta$ was done using a polarization modulation technique, in which the relative phase of two orthogonal linear polarizations was modulated that pass through a photoelastic modulator (PEM) \cite{sato1981measurement}. The change in rotation $\theta$ and ellipticity $\eta$ were probed simultaneously (see SI~\cite{SI}).\\
%Magnetization measurements have been performed using a superconducting quantum interference device (SQUID, Quantum Design, MPMS) in magnetic fields applied within the $ab$-planes at a temperature of $3$~K. The magnetic field was swept from 0 to 0.5 T in steps of 100 mT, while the step size was reduced to 50 mT for the field regime of 0.5 to 2.5 to resolve the anharmonicities near the low-field spin-flop transition in the MO responses. At each field, 3 measurements of 5 cycles each have been performed from which the magnetization and the standard deviation have been extracted.\\

\textbf{Data Availability}\\
The data supporting the findings of this study are available from the corresponding authors on reasonable request.\\

\textbf{Acknowledgments}\\
We thank Kevin Jenni for his kind assistance with the Laue measurements and valuable advice during the Laue data analysis. The authors acknowledge financial support funded by the Deutsche Forschungsgemeinschaft (DFG) through project No. 277146847-CRC1238, Control and Dynamics of Quantum Materials (subproject No. B05). AL and VT acknowledge support by the Deutsche Forschungsgemeinschaft (DFG) through the Transregional Research Collaboration TRR 80: Fom Electronic Correlations to Functionality (Augsburg, Munich, and Stuttgart). VT acknowledges the support via the project ANCD 20.80009.5007.19 (Moldova).\\

\textbf{Author contributions}\\
J.W. and H.H. performed the magneto-optical measurements and analyzed the data. P.v.L. and H.H. supervised the study. J.W. developed the theoretical model with the help of D.I.K. L.W. carried out the SQUID measurements. V.T. and A.L. synthesized the high-quality samples. J.W. and H.H. wrote the manuscript. J.W., A.S., R.V., L.W., Z.W., V.T., A.L., D.I.K., H.H., and P.v.L. discussed the results and commented on the paper.\\

\textbf{Competing interests}\\
The authors declare no competing financial or non-financial interests.

\def\bibsection{\section*{~\refname}} 
\bibliographystyle{naturemag}
\bibliography{bibliography}

%\section{Acknowledgments\label{ack}}
%We thank Kevin Jenni for his kind assistance with the Laue measurements and valuable advice during the Laue data analysis. %We kindly acknowledge XYZ for helpful discussions and valuable suggestions.
%The authors acknowledge financial support funded by the Deutsche Forschungsgemeinschaft (DFG) through project No. 277146847-CRC1238, Control and Dynamics of Quantum Materials (subproject No. B05). \textcolor{red}{Financial support AUGSBURG?!}

%\section{Additional information}
%\textbf{Supplementary information} is available for this paper at https://doi.org/xxx\\
%\textbf{Correspondence} and requests for materials should be addressed to P.v.L. or H.H.\\

\end{document}